\documentclass[conference]{IEEEtran}
\IEEEoverridecommandlockouts
\usepackage{cite}
\usepackage{graphicx}
\usepackage{textcomp}
\usepackage{xcolor}
\usepackage{hyperref,stfloats,cuted}
\usepackage[acronym,shortcuts]{glossaries}
\usepackage{amsmath, bm, amssymb, algpseudocode, algorithm, mathtools, svg,ifluatex, pgfplots, amsthm, caption, subcaption,bookmark}

\pgfplotsset{compat=1.15}

\newcommand{\ar}{\bm{a}_{R}}
\newcommand{\at}{\bm{a}_{T}}
\newcommand{\CC}{\mathbb{C}}

\newcommand{\expc}{\mathbb{E}}
\newcommand{\tr}{\text{Tr}}
\newcommand{\herm}{^\text{H}}
\newcommand{\vect}{\text{vec}}
\newcommand{\pard}[2]{\frac{\partial {#1}}{\partial {#2}}}
\newcommand{\br}[1]{\left( #1 \right)}
\newcommand{\brsq}[1]{\left[ #1 \right]}
\newcommand{\brcur}[1]{\left\{ #1 \right\}}
\newcommand{\re}[1]{\text{Re}\left[ #1 \right]}

\newcommand{\fim}{\bm{\mathcal{I}}}

\allowdisplaybreaks

\newacronym{mimo}{MIMO}{multi-input-multi-output}
\newacronym{ofdm}{OFDM}{orthogonal frequency-division multiplexing}
\newacronym{5g}{5G}{fifth generation}
\newacronym{6g}{6G}{sixth generation}
\newacronym{b5g}{B5G}{Beyond 5G}
\newcommand{\mimoofdm}{\ac{mimo}-\ac{ofdm} }
\newacronym{crlb}{CRB}{Cram\'er-Rao bound}
\newacronym{fim}{FIM}{Fisher information matrix}
\newacronym{repms}{REPMS}{Riemannian Exact Penalty Method via Smoothing}
\newacronym{srepms}{SREPMS}{Stochastic Riemannian Exact Penalty Method via Smoothing}
\newacronym{sdr}{SDR}{semidefinite relaxation}
\newacronym{sdp}{SDP}{semidefinite programming}
\newacronym{rcg}{RCG}{Riemannian conjugate gradient}
\newacronym{mmwave}{mmWave}{millimeter wave}
\newacronym{isac}{ISAC}{integrated sensing and communication}
\newacronym{aoa}{AoA}{angle of arrival}
\newacronym{aod}{AoD}{angle of departure}
\newacronym{qos}{QoS}{Quality of Service}
\newacronym{th}{THz}{Terahertz}
\newacronym{dof}{DoFs}{degrees of freedom}
\newacronym{iot}{IoT}{Internet of Things}
\newacronym{uav}{UAV}{unmanned aerial vehicle}
\newacronym{tx}{Tx}{transmitter}
\newacronym{rx}{Rx}{receiver}
\newacronym{ula}{ULA}{uniform linear array}
\newacronym{dft}{DFT}{Discrete Fourier Transform}
\newacronym{re}{RE}{resource element}
\newacronym{dl}{DL}{Downlink}
\newacronym{ul}{UL}{Uplink}
\newacronym{psd}{PSD}{positive semidefinite}
\newacronym{mse}{MSE}{mean squared error}
\newacronym{svd}{SVD}{singular value decomposition}
\newacronym{cpu}{CPU}{central processing unit}
\glsdisablehyper

\makeatletter
\def\ps@IEEEtitlepagestyle{%
  \def\@oddfoot{\mycopyrightnotice}%
  \def\@oddhead{\hbox{}\@IEEEheaderstyle\leftmark\hfil\thepage}\relax
  \def\@evenhead{\@IEEEheaderstyle\thepage\hfil\leftmark\hbox{}}\relax
  \def\@evenfoot{}%
}
\def\mycopyrightnotice{%
  \begin{minipage}{\textwidth}
  \centering \scriptsize
  Copyright~\copyright~20XX IEEE.  Personal use of this material is permitted.  Permission from IEEE must be obtained for all other uses, in any current or future media, including reprinting/republishing this material for advertising or promotional purposes, creating new collective works, for resale or redistribution to servers or lists, or reuse of any copyrighted component of this work in other works.
  \end{minipage}
}
\makeatother

\begin{document}
\title{Optimal and Robust Waveform Design for MIMO-OFDM Channel Sensing: A Cram\'er-Rao Bound Perspective}

\author{\IEEEauthorblockN{Xinyang Li\IEEEauthorrefmark{1}\IEEEauthorrefmark{2},
Vlad C. Andrei\IEEEauthorrefmark{1}\IEEEauthorrefmark{3}, Ullrich J. M\"onich\IEEEauthorrefmark{1}\IEEEauthorrefmark{4} and
Holger Boche\IEEEauthorrefmark{1}\IEEEauthorrefmark{5}}\thanks{
    The authors were supported in part by the
    German Federal Ministry of Education and Research (BMBF)
    in the programme “Souver\"an. Digital. Vernetzt.”
    within the research hub 6G-life under Grant 16KISK002.
    U. M\"onich and H. Boche were supported in part by the BMBF within the project "Post Shannon Communication - NewCom" under Grant 16KIS1003K.
    }
\IEEEauthorblockA{\IEEEauthorrefmark{1}Chair of Theoretical Information Technology, Technical University of Munich, Munich, Germany\\
\IEEEauthorrefmark{1}BMBF Research Hub 6G-life,
\IEEEauthorrefmark{5}Munich Center for Quantum Science and Technology,
\IEEEauthorrefmark{5}Munich Quantum Valley\\
Email: \IEEEauthorrefmark{2}xinyang.li@tum.de,
\IEEEauthorrefmark{3}vlad.andrei@tum.de,
\IEEEauthorrefmark{4}moenich@tum.de,
\IEEEauthorrefmark{5}boche@tum.de}
}

\maketitle

\begin{abstract}
Wireless channel sensing is one of the key enablers for \ac{isac} which helps communication networks understand the surrounding environment. In this work, we consider MIMO-OFDM systems and aim to design optimal and robust waveforms for accurate channel parameter estimation given allocated OFDM resources. The \ac{fim} is derived first, and the waveform design problem is formulated by maximizing the log determinant of the FIM. We then consider the uncertainty in the parameters and state the stochastic optimization problem for a robust design. We propose the \ac{repms} and its stochastic version SREPMS to solve the constrained non-convex problems. In simulations, we show that the \ac{repms} yields comparable results to the \ac{sdr} but with a much shorter running time. Finally, the designed robust waveforms using SREMPS are investigated, and are shown to have a good performance under channel perturbations.
\end{abstract}


\glsresetall

\section{Introduction}\label{sec:intro}
Since the deployment and evolution of the \ac{5g} communications technology, the reliability and connectivity of wireless systems have been improved impressively. Thanks to the large bandwidth at high radio frequencies\cite{heath2016overview} and the \ac{ofdm} techniques, current networks can provide enormous data rates and serve massive number of users at the same time. \Ac{mimo} techniques, on the other hand, are used to take advantage of additional spatial \ac{dof} so that the use scenarios are further expanded.

In future communication systems (\ac{b5g} and 6G)\cite{saad2019vision}, the functionalities of intelligence and perception are expected to be introduced in order to enhance \ac{qos} and support more advanced and complicated applications such as the \ac{iot}\cite{he2022collaborative} and \ac{uav} networks\cite{zeng2019accessing}. As a key enabler, \ac{isac}\cite{liu2022integrated} opens the eyes of future wireless systems, in which the communication partners can sense the surrounding environment with the aid of received radio signals. Due to the sparse signal propagation paths at short wavelengths, wireless channels carry a large amount of environmental information and are an important feature that can be exploited. Compared to the conventional channel models that assume rich scattering environments, the beam-space channel model\cite{heath2016overview, zhang2021overview} leverages the propagation geometric structure and is determined by a set of multipath parameters, i.e., path gain, path delay, Doppler shift, \ac{aoa} and \ac{aod}. 

Designing appropriate waveforms is an important step in \ac{isac} systems, as it determines the performance limits the systems can achieve. Different design criteria can be chosen\cite{liu2022survey} to meet different requirements. For example, the authors of \cite{liu2018mu} aim to design the beamformer by matching the radar beampattern while satisfying the SINR constraints. For parameter estimation, a well-known performance bound is the \ac{crlb}\cite{kay1993fundamentals}, which states a lower bound on the mean squared estimation error of any unbiased estimator. Waveform design by optimizing \ac{crlb} is studied in\cite{li2007range,ni2020waveform,liyanaarachchi2021optimized}, but they only focus on a subset of multipath parameters and rarely consider the \ac{mimo}-\ac{ofdm} case. In addition, the robustness of the designed waveforms to the changing environment is another important criterion in practice but lacks formulation and analysis in related works. Moreover, the formulated optimization problems of waveform design are generally non-convex. Classical convex relaxation approaches, such as \ac{sdr}\cite{luo2010semidefinite}, suffer from high computational complexity, especially for stochastic optimization. Riemannian manifold methods\cite{liu2018mu}, on the other hand, converge faster than \ac{sdr}, but require further investigation and appropriate algorithms when additional constraints are present.

In this work, we derive the \ac{fim}, the inverse of \ac{crlb}, of multipath parameters and formulate the optimization problems to design optimal and robust sensing waveforms compatible with current \mimoofdm communication systems. We adopt the \ac{repms}\cite{liu2020simple} and develop its stochastic version \ac{srepms} to solve the problems. In numerical experiments, the running time of the proposed algorithms and the performance of resulting waveforms are analyzed.

\section{System Model}

Considering a \mimoofdm communication system comprising a \ac{tx} and a \ac{rx}, each of which is equipped with a half wavelength spacing \ac{ula} with $N_T$ and $N_R$ antenna elements respectively, the frequency representation of the channel matrix\cite{zhang2021overview} consisting of $L$ paths at subcarrier $n$ and \ac{ofdm} symbol $k$ is
\begin{equation}
    \bm{H}_{n,k} = \sum_{l=1}^Lb_l\omega_{n,k,l}\ar(\phi_l)\at(\theta_l)^\top,
    \label{ofdmchannel}
\end{equation}
with $\omega_{n,k,l} = e^{-j2\pi nf_0 \tau_l}e^{j2\pi f_{D,l}kT_s}$, where
\begin{itemize}
    \item $f_0$ is the \ac{ofdm} subcarrier spacing,
    \item $T_s$ is the \ac{ofdm} symbol duration,
    \item $b_l \in \CC$ is the channel gain of the $l$-th path,
    \item $\tau_l$ is the $l$-th path delay,
    \item $f_{D,l}$ is the Doppler shift of the $l$-th path,
    \item $\at(\theta_l) \in \CC^{N_T}$ is the \ac{ula} response vector at \ac{aod} $\theta_l$,
    \item $\ar(\phi_l)\in \CC^{N_R}$ is the \ac{ula} response vector at \ac{aoa} $\phi_l$.
\end{itemize}

Suppose the \ac{tx} transmits signal $\bm{x}_{n,k}$ at the $(n,k)$-th \ac{ofdm} \ac{re}, then the received signal is
\begin{equation}
\bm{y}_{n,k} = \bm{H}_{n,k}\bm{x}_{n,k} + \bm{z}_{n,k},
\label{sysmodelinfreq}
\end{equation}
with $\bm{z}_{n,k}$ being the additive Gaussian noise of zero mean and covariance matrix $\bm{C}_{\bm{z}_{n,k}}=\sigma_{n,k}^2\bm{I}$. The noise vectors on different \acp{re} are assumed to be independent.

Compared to the classical channel estimation, where only the channel matrices $\bm{H}_{n,k}$ are of interest, in future networks it will be vital to additionally extract their multipath information \cite{liu2022integrated}, i.e., $\{b_l, \tau_l, f_{D,l}, \phi_l, \theta_l\}$ for all $l$. To this end, the waveforms $\bm{x}_{n,k}$ should be designed carefully for accurate parameter estimation. 
In \ac{ofdm}, the sensing waveforms can either be designed jointly with the communication symbols, or allocated to the dedicated \acp{re} and designed independently of the communication symbols. In this work, we focus on the latter case, as its implementation is compatible with current standards such as 5G NR\cite{liyanaarachchi2021optimized}, where the \acp{re} are not always fully occupied.


\subsection{Cram\'er-Rao Bound (CRB)}
Given the observation $\bm{y}$ that depends on the parameter $\bm{\xi}$ to be estimated and the conditioned distribution $p(\bm{y}|\bm{\xi})$, the \ac{mse} of any unbiased estimator $\hat{\bm{\xi}}(\bm{y})$ is bounded by the inverse of the \ac{fim} $\fim$, i.e., $\expc[(\hat{\bm{\xi}}(\bm{y})-\bm{\xi})(\hat{\bm{\xi}}(\bm{y})-\bm{\xi})\herm] \succeq \fim^{-1}$ and $\bm{P}\succeq \bm{Q}$ indicates that $\bm{P} - \bm{Q}$ is a \ac{psd} matrix\cite{kay1993fundamentals,larsen2009performance} for two \ac{psd} matrices $\bm{P}$ and $\bm{Q}$. When $p(\bm{y}|\bm{\xi})$ is a complex Gaussian distribution\cite{kay1993fundamentals}, the value at the $i$-th row and $j$-th column of \ac{fim} is
\begin{equation}
    \begin{split}
        [\fim]_{i,j} = &\tr\brsq{\bm{C}_{\bm{y}}^{-1}(\bm{\xi})\frac{\partial \bm{C}_{\bm{y}}(\bm{\xi})}{\partial \xi_i}\bm{C}_{\bm{y}}^{-1}(\bm{\xi})\frac{\partial \bm{C}_{\bm{y}}(\bm{\xi})}{\partial \xi_j}}\\
        &+ 2\text{Re} \left[\frac{\partial \bm{\mu}\herm(\bm{\xi})}{\partial \xi_i}\bm{C}_{\bm{y}}^{-1}(\bm{\xi})\frac{\partial \bm{\mu}(\bm{\xi})}{\partial \xi_j}\right],
    \end{split}
    \label{gaussfim}
    \end{equation}
where $\xi_i$ denotes the $i$-th component of $\bm{\xi}$, and $\bm{\mu}(\bm{\xi})$, $\bm{C}_{\bm{y}}(\bm{\xi})$ are the corresponding mean and covariance matrix of $\bm{y}$ dependent on $\bm{\xi}$. $\tr\brsq{\cdot}$ and $\re{\cdot}$ indicate the trace and real part of a complex matrix respectively. Rewriting \eqref{sysmodelinfreq} using the vectorization formula $\vect(\bm{ABC})=(\bm{C}^\top \otimes \bm{A})\vect(\bm{B})$ with the Kronecker product $\otimes$ results in
\begin{equation}
\bm{y}_{n,k} = (\bm{x}_{n,k}^\top \otimes \bm{I}) \bm{h}_{n,k} + \bm{z}_{n,k},
\end{equation}
in which $\bm{I}$ is the identity matrix and
\begin{equation}
    \bm{h}_{n,k} = \vect\br{\bm{H}_{n,k}} = \sum_{l=1}^L b_l \omega_{n,k,l} \at(\theta_l) \otimes \ar(\phi_l).
\end{equation}

We assume that the multipath parameters are independent on $(n,k)$. Without loss of generality, we allocate sensing symbols on $M$ \ac{ofdm} REs, say, $\{(n_m,k_m)\}_{m=1}^M$. In the following, we use subscript $m$ to indicate $(n_m,k_m)$ for convenience. To express the dependence of the channel on the parameters explicitly, we write $\bm{h}_m$ as $\bm{h}_m(\bm{\xi})$, with $\bm{\xi}=\text{vec}\br{[\bm{\xi}_1, \bm{\xi}_2, ..., \bm{\xi}_L]^\top}$ collecting all multipath parameters,
\begin{equation}
    \bm{\xi}_l = \begin{bmatrix}
        b_{l,R} &b_{l,I} &\tau_l &f_{D,l} &\phi_l &\theta_l
        \end{bmatrix}^\top,
\label{param}
\end{equation}
and $b_{l,R}$ and $b_{l,I}$ denote the real and imaginary part $b_l$, respectively. Therefore, \eqref{gaussfim} can be rewritten as
\begin{equation}
    [\fim]_{i,j} = \sum_{m=1}^M \frac{2}{\sigma_m^2}\text{Re}\left[\left(\frac{\partial \bm{h}_m(\bm{\xi})}{\partial \xi_i}\right)\herm(\bm{x}_m^*\bm{x}_m^\top \otimes \bm{I})\frac{\partial \bm{h}_m(\bm{\xi})}{\partial \xi_j}\right],
\label{gaussfim3}
\end{equation}
where we use the mixed-product property of Kronecker product. The resulting FIM is derived in Appendix~\ref{crlbder} as
\begin{equation}
\fim = \re{\br{\sum_{m=1}^M \frac{2}{\sigma_m^2}\bm{\Lambda}_m\herm \bm{T}\herm \bm{x}_m^*\bm{x}_m^\top \bm{T\Lambda}_m}\circ \br{\bm{R}\herm \bm{R}}},
\label{eq:fim}
\end{equation}
where $\circ$ indicates the Hadamard product, and $\bm{\Lambda}_m$, $\bm{T}$, $\bm{R}$ depend on the parameters and are given in \eqref{eq:fimparam}.
\subsection{Optimal Design}\label{subsec:opt}
There are several ways to construct objective functions on $\fim$ to be optimized so that the \ac{crlb} is correspondingly minimized\cite{li2007range}. We choose to maximize the determinant of \ac{fim} since it controls the element scaling while calculating the matrix inverse. In addition, depending on different use cases, we need to place more importance on certain parameters, e.g., delays and \acp{aoa} are much more relevant for indoor localization\cite{wen2019survey}. To this end, we can multiply $\fim$ with a weighting matrix $\bm{J}$ from the right and its conjugate transpose $\bm{J}\herm$ from the left to scale the associated parts. By imposing the power constraints and defining $\bm{X} = [\bm{x}_1, \bm{x}_2, \dots, \bm{x}_M]$, the overall optimization problem can be expressed as
\begin{equation}
    \begin{split}
        &\max_{\bm{X}} \log\det\br{\bm{J}\herm\fim\bm{J}} \\
        \text{s.t.} \quad &\frac{1}{M} \tr{\brsq{\bm{XX}\herm}} \le P\\
        & \left\|\bm{x}_m\right\|_2^2 \le P_m, \quad \forall m=1\dots M.
    \end{split} \tag{P1} \label{P1}
\end{equation}
\eqref{P1} is a nonconvex problem and can be solved using the \ac{sdr} technique, but the number of variables in the relaxed problem increases quadratically with the number of transmit antennas $N_T$, resulting in a high computational load. To alleviate this problem, we observe that the total power is always exhausted to reach a higher objective value. Hence, the total power inequality constraint in \eqref{P1} can then be replaced by equality, which leads to the hypersphere manifold 
\begin{equation}
    \mathcal{S} = \left\{\bm{X} \in \mathbb{C}^{N_T\times M} \middle| \tr\brsq{\bm{XX}\herm} = \left\|\bm{X}\right\|_F^2 = MP\right\},
\end{equation}
and the resulting problem is reformulated as a constrained manifold optimization problem in the following:
\begin{equation}
    \begin{split}
    &\max_{\bm{X}\in \mathcal{S}} \log\det\br{\bm{J}\herm\fim\bm{J}} \\
    \text{s.t.} \quad & \left\|\bm{x}_m\right\|^2 \le P_m, \quad \forall m=1\dots M.
    \end{split}
    \tag{P1M}
    \label{P1M}
\end{equation}



\subsection{Robust Design}
It should be mentioned that the objective function in the optimal design problem depends on $\bm{\xi}$ through $\fim$. A full and precise knowledge of the parameters can result in optimal waveforms. However, in practice, such assumption is unrealistic, due to, for instance, the changing of environment, parameter estimation error and feedback delay. To this end, we assume that the \ac{tx} only knows the perturbed parameters $\hat{\bm{\xi}}$, while the true value $\bm{\xi} = \hat{\bm{\xi}} + \Delta \bm{\xi}$ is unavailable. The error $\Delta \bm{\xi}$ follows the Gaussian distribution $\mathcal{N}(\bm{0}, \bm{C}_e)$, with a diagonal covariance matrix $\bm{C}_e$ comprising diagonal elements $\sigma_{b_{l,R}}^2, \sigma_{b_{l,I}}^2,\sigma_{\tau_l}^2,\sigma_{f_{D,l}}^2,\sigma_{\phi_l}^2,\sigma_{\theta_l}^2$ for all $l$ corresponding to the error variances of respective parameters. The unknown true parameter $\bm{\xi}$ thus follows $\mathcal{N}(\hat{\bm{\xi}}, \bm{C}_e)$. In the following, we also take into account the uncertainty in the error variances and assume they are randomly distributed.

From the perspective of robust optimization\cite{beyer2007robust}, one option is to optimize over the expectation of the objective function. Thus, the robust design problem can be formulated as
\begin{equation}
    \begin{split}
    &\max_{\bm{X} \in \mathcal{S}} \mathbb{E}\brsq{\log\det\br{\bm{J}\herm\fim\bm{J}}}\\
    \text{s.t.} \quad & \left\|\bm{x}_m\right\|^2 \le P_m, \quad \forall m=1\dots M,
    \end{split}
    \tag{P1E}
    \label{P1E}
\end{equation}
with $\mathbb{E}$ taken over $\bm{C}_e$ and $\bm{\xi}$. One issue to solve \eqref{P1E} is that it's intractable to derive the closed form of the objective function. A common approach is to apply the stochastic method by sampling points randomly and calculating the empirical mean value as an approximation, which will be discussed in \ref{sec:sop}.

\section{Optimization}
\subsection{Semidefinite Relaxation}

We introduce the new variables $\bm{R}_m=\bm{x}_m^*\bm{x}_m^\top$ for all $m$, relax its rank 1 constraint and the optimization problem \eqref{P1} becomes
\begin{equation}
\begin{split}
    &\max_{\bm{R}_1,\bm{R}_2, ..., \bm{R}_M} \log\det\br{\bm{J}\herm\fim\bm{J}} \\
    \mathrm{s.t.} \quad &\frac{1}{M}\tr\brsq{\sum_{m=1}^M\bm{R}_m} \le P\\
    &\bm{R}_m\succeq \bm{0}, \quad \tr\brsq{\bm{R}_m}\le P_m, \forall m = 1,...,M,
\end{split}
\tag{P1R}
\label{P1R}
\end{equation}
which is shown as \ac{sdp}. After solving \eqref{P1R}, one can approximate the rank 1 results by \ac{svd} or randomization\cite{luo2010semidefinite}.

\subsection{Manifold Optimization}\label{sec:manopt}


Due to the additional constraints on the symbol norm, the traditional unconstrained methods like \ac{rcg}\cite{liu2018mu} cannot be directly applied. We therefore modify the problem by adding the constraint as a penalty term to the objective function. Specifically, we employ the \ac{repms}\cite{liu2020simple}, in which the problem \eqref{P1M} can be reformulated as 
\begin{equation}
    \begin{split}
        &\min_{\bm{X}\in\mathcal{S}} \mathcal{L}(\bm{X}, \rho, u) =\\
        &\qquad -\log\det\br{\bm{J}\herm\fim\bm{J}} + \rho\sum_{m=1}^M p_u\br{\|\bm{x}_m\|_2^2-P_m},
    \end{split}
\end{equation}
with $\rho$, $u$ being the penalty weight, smoothing factor respectively, and the linear-quadratic loss is given by
\begin{equation}
    p_u(x) = \begin{cases}
        0 & x\le 0\\
        \frac{x^2}{2u} & 0\le x\le u\\
        x-\frac{u}{2} & x\ge u
    \end{cases}.
\end{equation}

Choosing \ac{rcg} as the base solver, we end up with the \ac{repms}\cite{liu2020simple} in Algorithm~\ref{alg:repms} for optimal waveform design. At the $k$-th step, the update direction $\bm{p}_k$ is computed based on the previous direction and the current Riemannian gradient $\mathrm{grad}\mathcal{L}$ through the function $\tau$, which can be chosen according to different rules\cite{shewchuk1994introduction}. The retraction $\gamma$ associated to the hypersphere manifold\cite{liu2018mu} is used as the update function that moves the point $\bm{X}_k$ along $\bm{p}_k$ and keeps it on $\mathcal{S}$. 

\begin{algorithm}
    \caption{Optimal waveform design using \ac{repms}}
    \begin{algorithmic}
    \Require Function $\mathcal{L}$, initial point $\bm{X}_{0}\in \mathcal{S}$, initial penalty weight $\rho_0$, $\theta_{\rho}>1$, $\rho_{\mathrm{max}}$, initial smoothing factor $u_0$, $0<\theta_u<1$, $u_{\mathrm{min}}$
    \Ensure Optimal $\bm{X}$
    \State $k \gets 0$
    \State $\bm{p}_0  \gets  -\mathrm{grad}\mathcal{L}(\bm{X}_0, \rho_0, u_0)$
    \While{Stopping criterion not met}
    \State Compute update step size $t_k$ by certain rules
    \State $\bm{X}_{k+1}\gets \gamma(\bm{X}_k, \bm{p}_k, t_k)$
    \State $\rho_{k+1} \gets \min\brcur{\theta_{\rho}\rho, \rho_{\mathrm{max}}}$
    \State $u_{k+1} \gets \max\brcur{\theta_{u}u, u_{\mathrm{min}}}$
    \State $\bm{p}_{k+1}  \gets \tau\br{\bm{p}_k, \mathrm{grad}\mathcal{L}(\bm{X}_{k+1}, \rho_{k+1}, u_{k+1})}$
    \EndWhile

    \State \Return $\bm{X}_k$
    \end{algorithmic}
    \label{alg:repms}
\end{algorithm}

\subsection{Stochastic Optimization}\label{sec:sop}
To solve \eqref{P1E}, we apply the idea of stochastic optimization, in which the expectation is approximated by the sample means using a Monte Carlo approach, namely
\begin{equation}
   \frac{1}{N}\sum_{n=1}^N \log\det\br{\bm{J}\herm\fim_n\bm{J}} \approx  \mathbb{E}\brsq{\log\det\br{\bm{J}\herm\fim\bm{J}}},
   \label{stoch}
\end{equation}
where $\fim_n$ is computed on $\bm{\xi}_n$ with $\bm{\xi}_n\sim \mathcal{N}(\hat{\bm{\xi}}, \bm{C}_{e,n})$ and the diagonal elements of $\bm{C}_{e,n}$ are sampled randomly from the predefined distributions. It should be mentioned that evaluating the sample mean brings additional computational complexity and \ac{sdr} might become infeasible in practice for large sample size $N$ thus will not be used to solve the robust design problem in our work.

Similar to Algorithm~\ref{alg:repms}, we reformulate \eqref{P1E} to an unconstrained case as
\begin{equation}
    \begin{split}
    &\min_{\bm{X}\in\mathcal{S}}\mathcal{SL}(\bm{X}, \brcur{\bm{\xi}_n}_{n=1}^N,\rho, u)=\\
    &\quad - \frac{1}{N}\sum_{n=1}^N \log\det\br{\bm{J}\herm\fim_n\bm{J}}  + \rho\sum_{m=1}^M p_u\br{\|\bm{x}_m\|_2^2-P_m}.
    \end{split}
\end{equation}
While optimizing, at each new iteration we sample a new set of $N$ parameter vectors to compute the sample mean and apply one step \ac{repms}. Finally, we summarize the \ac{srepms} for robust waveform design in Algorithm~\ref{alg:SPRM}.
\begin{algorithm}
\caption{Robust waveform design using \ac{srepms}}
\begin{algorithmic}
\Require 
Function $\mathcal{SL}$, $N$, $\hat{\bm{\xi}}$, distributions of error variances, $\bm{J}$, $\bm{X}_{0}\in \mathcal{S}$, $\rho_0$, $\theta_{\rho}>1$, $\rho_{\mathrm{max}}$, $u_0$, $0<\theta_u<1$, $u_{\mathrm{min}}$
\Ensure Robust $\bm{X}$
\State Sample $\brcur{\bm{C}_{e,n}}_{n=1}^N$ from the given distributions
\State Sample $\bm{\xi}_n$ from $\mathcal{N}(\hat{\bm{\xi}}, \bm{C}_{e,n})$  for $n=1...N$
\State $\bm{p}_0 \gets -\mathrm{grad} \mathcal{SL}(\bm{X}_0, \brcur{\bm{\xi}_n}_{n=1}^N,\rho_0, u_0)$
\State $k \gets 0$
\While{Stopping criterion not met}
\State Compute update step size $t_k$
\State $\bm{X}_{k+1} \gets \gamma(\bm{X}_{k}, \bm{p}_k, t_k)$
\State Sample $\brcur{\bm{C}_{e,n}}_{n=1}^N$ from the given distributions
\State Sample $\bm{\xi}_n$ from $\mathcal{N}(\hat{\bm{\xi}}, \bm{C}_{e,n})$  for $n=1...N$
\State $\rho_{k+1} \gets \min\brcur{\theta_{\rho}\rho, \rho_{\mathrm{max}}}$
\State $u_{k+1} \gets \max\brcur{\theta_{u}u, u_{\mathrm{min}}}$
\State $\bm{p}_{k+1}  \gets \tau\br{\bm{p}_k, \mathrm{grad}\mathcal{SL}(\bm{X}_{k+1}, \rho_{k+1}, u_{k+1})}$
\State $k \gets k+1$
\EndWhile
\State \Return $\bm{X}_k$
\end{algorithmic}
\label{alg:SPRM}
\end{algorithm}

\section{Numerical Results}
In the simulations, we consider an $8\times 8$ \ac{mimo} system and set the carrier frequency $f_c$ to 3 GHz and the subcarrier spacing $f_0$ to 15 kHz. We allocate the sensing waveforms in a rectangular \ac{ofdm} region, occupying 128 subcarriers and 14 \ac{ofdm} symbols (one slot in 5G). The average transmit power $P$ and SNR are fixed to $10$ and $-10$ dB, respectively. The symbol-wise power constraint $P_m$ is set to $\alpha P$, where $\alpha>1$ for all $m$. The scaling matrix $\bm{J}$ is a diagonal matrix with the diagonal elements $T_s=\frac{1}{f_0}$ and $f_0$ for the path delay and Doppler shift parts, respectively, and 1 otherwise, to avoid the numerical instability caused by the large difference in the orders of the values in $\fim$. The number of channel paths is set to 3, and we generate 100 realizations of multipath parameters randomly according to the following settings:
\begin{itemize}
    \item The real and imaginary parts of path gains are from zero mean unit variance normal distribution,
    \item By sampling the path lengths from a uniform distribution between 10 and 800 meters, the path delays are computed,
    \item By sampling the relative velocities of different paths from a uniform distribution between 0 and 80 m/s, the Doppler shifts are computed,
    \item AoAs and AoDs are sampled uniformly between $-90^\circ$ and $90^\circ$.
\end{itemize}
In the following, the optimization tool MOSEK\cite{mosek} is used for \ac{sdr} while \ac{repms} and \ac{srepms} are implemented with Pymanopt\cite{JMLR:v17:16-177}.

\begin{figure}[h]
    \centering
    \includegraphics[width=.48\textwidth]{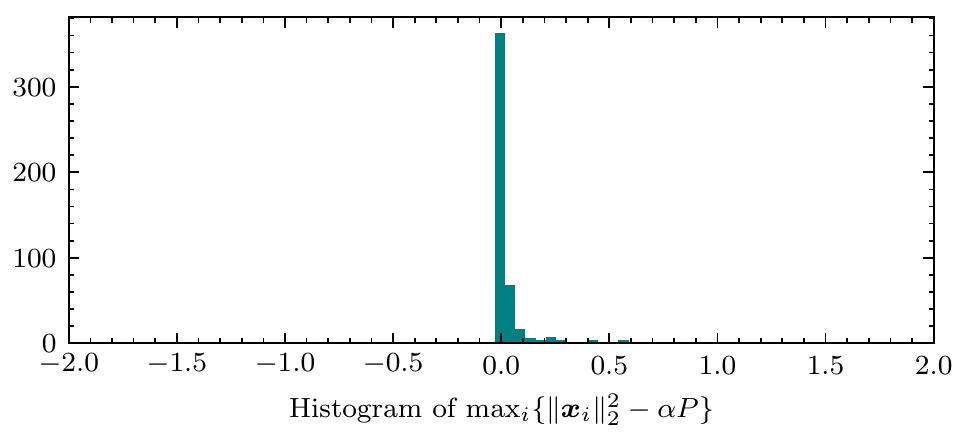}
    \caption{Histogram of the maximum symbol norm resulting from \ac{repms} subtracted by the threshold.}
    \label{fig:gihist}
\end{figure}
\begin{figure}[h]
    \centering
    \includegraphics[width = .48\textwidth]{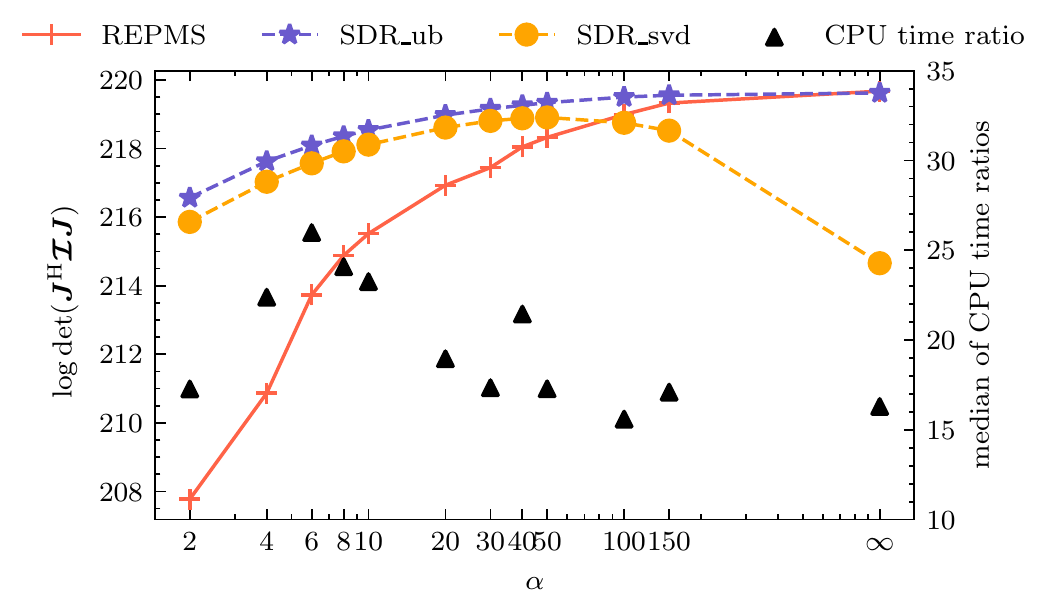}
    \caption{Comparison of optimization results and the \acs{cpu} time ratios (\ac{sdr} to \ac{repms}). \textbf{ub} (upper bound) and \textbf{svd} indicate the \ac{sdr} solutions without and with \ac{svd} rank 1 recovery.}
    \label{sdrvsmanopt}
\end{figure}
\begin{figure}[h]
    \centering
    \includegraphics[width=.48\textwidth]{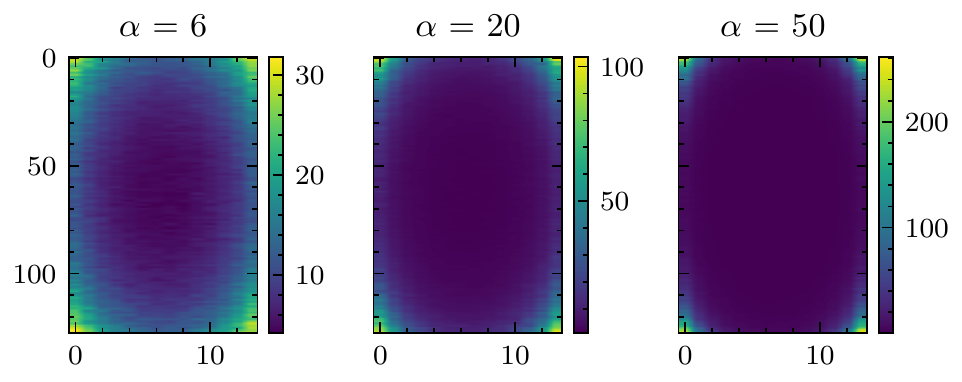}
    \caption{Resulting average power allocation of \ac{repms}.}
    \label{fig:ofdmwo}
\end{figure}

We first investigate the feasibility of the \ac{repms} results. The associated parameters are $\rho_0=1$, $u_0=1$,  $u_{\text{min}}=10^{-6}$ and $\theta_u = \br{u_{\text{min}}/u_0}^{\frac{1}{30}}$ by following the experiment settings in\cite{liu2020simple}. $\rho_{\text{max}}$ should be as large as possible but the numerical overflow need to be avoided thus is set to $2^{20}$. We also choose $\theta_{\rho} = 2$ as it yields a faster convergence in simulations. We apply Algorithm~\ref{alg:repms} to the generated channel realizations and plot the histogram of the differences between the resulting maximum symbol power with the threshold $\alpha P$, i.e., $\max_i\brcur{\|\bm{x}_i\|_2^2-\alpha P}$, for $\alpha = 2, 4, 6, 8, 10$ in Fig.~\ref{fig:gihist}. It turns out that \ac{repms} can lead to feasible solutions when a small tolerance allowed.

To compare the performance between \ac{sdr} and \ac{repms}, the averaged optimized objective values and the medians of consumed \ac{cpu} time ratios (\ac{sdr} to \ac{repms}) are plotted in Fig.~\ref{sdrvsmanopt} for varying $\alpha$ ($\infty$ means no symbol-wise constraint). It's obvious that the \ac{sdr} method produces larger objective values for low values of $\alpha$. After a certain threshold, the \ac{repms} outperforms the \ac{sdr} with SVD recovery and even reaches the \ac{sdr} upper bound despite the fact that the \ac{repms} algorithm can achieve more than $15$ times speedup than \ac{sdr} in terms of \ac{cpu} time on the same hardware platform, which makes the manifold optimization more attractive in our works. We also demonstrate the average power allocation on the given \ac{ofdm} resources in Fig.~\ref{fig:ofdmwo}, and notice that most power is concentrated on few \acp{re} at the grid corners, so the designed sensing waveforms are expected to have little impact on the communication resources.
\begin{figure}[h]
    \centering
    \includegraphics[width=.48\textwidth]{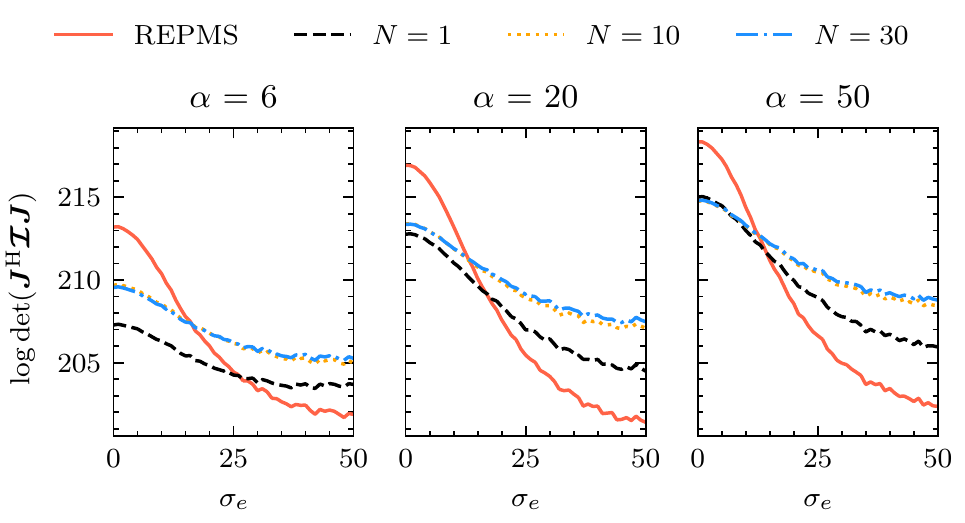}
    \caption{Performance of designed robust waveforms using \ac{srepms}.}
    \label{robustwf}
\end{figure}

To test the performance of the designed robust waveforms, we use a single parameter $\sigma_e$ to adjust the variances of all multipath parameters. In particular, we set the parameter perturbation variances as $\sigma_{b_{l,R}}=\sigma_{b_{l,I}}=10^{-2}\sigma_e$, $\sigma_{\tau_l}=10^{-8}\sigma_e$, $\sigma_{f_{D,l}}=5\sigma_e$, $\sigma_{\phi_l}=\sigma_{\theta_l}=10^{-2}\sigma_e$ for all $l$ and $\sigma_e$ is uniformly distributed from $0$ to $50$ ($\sigma_e=0$ means no perturbation). We apply Algorithm~\ref{alg:SPRM} by treating the previously generated 100 parameters as $\hat{\bm{\xi}}$ and set $N$ to 1, 10 and 30. To test the performance of different waveforms, for each fixed $\hat{\bm{\xi}}$ and $\sigma_e$, we generate another 100 random parameters as the true $\bm{\xi}$ from $\mathcal{N}(\hat{\bm{\xi}}, \bm{C}_e)$. The resulting average objective values against $\sigma_e$ are plotted in Fig.~\ref{robustwf}. It shows that as $N$ increases, the designed waveforms yield better performance, but at the cost of efficiency, as expected. We also observe a trade-off between the designed optimal and robust waveforms for different levels of uncertainty. Furthermore, the feasibility and power allocation of the \ac{srepms} results are similar to \ac{repms}. Due to the page limitation, we don't demonstrate them here.
\begin{figure}[h]
    \centering
    \includegraphics[width = .48\textwidth]{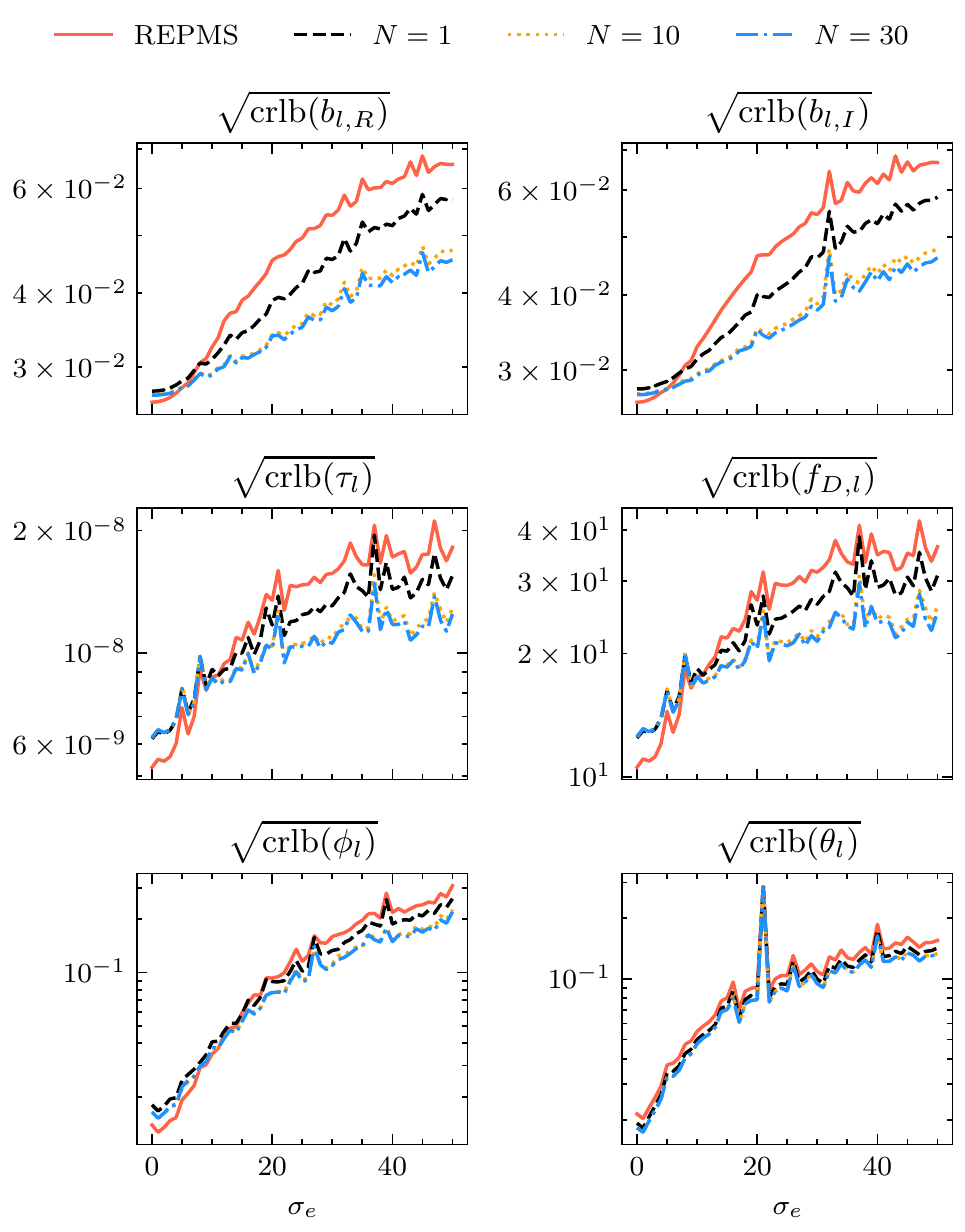}
    \caption{\ac{crlb} of multipath parameters resulting from previously designed waveforms at different perturbation levels.}
    \label{crlbval}
\end{figure}

Finally, we plot the resulting square root of the \ac{crlb} for each parameter against varying channel perturbation levels $\sigma_e$ in Fig.~\ref{crlbval} for $\alpha=50$. For the same type of parameter, we take their average over all paths. The nonsmooth curves reflect that a larger $\log\det$ of \ac{fim} can't always guarantee a lower \ac{crlb} on every parameter, but the overall results are shown to be consistent with the previous observations.

\section{Conclusion}
In this paper, we derive the \ac{crlb} for the multipath parameter estimation in \mimoofdm channel sensing. We formulate the waveform design problems to maximize the $\log\det$ of \ac{fim} under total and symbol-wise power constraints and introduce \ac{repms} to design optimal waveforms. By considering the uncertainty of parameters, the stochastic method \ac{srepms} for robust waveform design is then proposed. In numerical simulations, it shows that the manifold techniques can speed up the solving process over 15 times in terms of \ac{cpu} time and can still provide comparable results to \ac{sdr}. Finally, the performance trade-off of robust waveforms is demonstrated and the resulting \ac{crlb} for each parameter is presented.

\appendices
\section{}\label{crlbder}
The derivative of $\bm{h}_m(\bm{\xi})$ to each parameter are given as
\begin{align*}
\pard{\bm{h}_m(\bm{\xi})}{b_{l,R}}&=\omega_{m,l} \at(\theta_l) \otimes \ar(\phi_l)\\
\pard{\bm{h}_m(\bm{\xi})}{b_{l,I}}&=j\omega_{m,l} \at(\theta_l) \otimes \ar(\phi_l)\\
\pard{\bm{h}_m(\bm{\xi})}{\tau_l}&=b_l g_{m,l} \at(\theta_l) \otimes \ar(\phi_l)\\
\pard{\bm{h}_m(\bm{\xi})}{f_{D,l}}&=b_l f_{m,l} \at(\theta_l) \otimes \ar(\phi_l)\\
\pard{\bm{h}_m(\bm{\xi})}{\phi_l} &= b_l \omega_{m,l} \at(\theta_l) \otimes \bm{d}_R(\phi_l)\\
\pard{\bm{h}_m(\bm{\xi})}{\theta_l} &= b_l \omega_{m,l} \bm{d}_T(\theta_l) \otimes \ar(\phi_l)
\end{align*}
with $g_{m,l} = -j2\pi n_mf_0 \omega_{m,l}$, $f_{m,l} = j2\pi k_mT_s \omega_{m,l}$, $\bm{d}_R(\phi_l) =\pard{\ar(\phi_l)}{\phi_l}$ and $\bm{d}_T(\theta_l) = \pard{\at(\theta_l)}{\theta_l}$. It's then straightforward to show that $\pard{\bm{h}_m(\bm{\xi})}{\bm{\xi}} = \bm{T} * \bm{R\Lambda}_m$ where $*$ is the Kahtri-Rao product (column-wise Kronecker product), and the respective matrices are given in \eqref{eq:fimparam}. We define $\bullet$ as the face-splitting product (row-wise Kronecker product). Given matrices $\bm{A, B, C, D, E, F}$ we have the following properties:
\begin{itemize}
    \item $(\bm{A} * \bm{B})\herm = \bm{A}\herm \bullet \bm{B}\herm$,
    \item $(\bm{A} \bullet \bm{B})(\bm{C}\otimes \bm{D})(\bm{E} * \bm{F}) = (\bm{ACE})\circ (\bm{BDF})$,
    \item $\bm{A}\circ (\bm{BDF}) = (\bm{BAF})\circ \bm{D}$ if $\bm{B}$ and $\bm{F}$ are diagonal matrices,
    \item $\bm{A} \circ \bm{B} + \bm{C} \circ \bm{B} = \br{\bm{A}+\bm{C}}\circ \bm{B}$.
\end{itemize}
With these properties the \ac{fim} can be derived as
\begin{equation}
\begin{split}
\fim &= \sum_{m=1}^M \frac{2}{\sigma_m^2} \re{\br{\bm{T} * \bm{R\Lambda}_m}\herm \br{\bm{x}_m^*\bm{x}_m^\top \otimes \bm{I}} \br{\bm{T} * \bm{R\Lambda}_m}}\\
&= \sum_{m=1}^M \frac{2}{\sigma_m^2} \re{\br{\bm{T}\herm \bm{x}_m^*\bm{x}^\top\bm{T}} \circ \br{\bm{\Lambda}_m\herm\bm{R}\herm \bm{R\Lambda}_m}}\\
&=\re{\sum_{m=1}^M \br{\frac{2}{\sigma_m^2}\bm{\Lambda}_m\herm \bm{T}\herm \bm{x}_m^*\bm{x}_m^\top \bm{T\Lambda}_m}\circ \br{\bm{R}\herm \bm{R}}}.
\end{split}
\end{equation}

\begin{figure*}[t]
    \begin{equation}
            \begin{split}
            \bm{\Lambda}_m &= \text{diag}\{\dots \omega_{m,l} \dots j\omega_{m,l} \dots b_l g_{m,l} \dots b_l f_{m,l} \dots b_l \omega_{m,l} \dots b_l \omega_{m,l} \dots\}, \quad l = 1\dots L, \\
            \bm{T} &= \begin{bmatrix}
                \bm{A}_T(\bm{\theta}) & \bm{A}_T(\bm{\theta}) & \bm{A}_T(\bm{\theta}) & \bm{A}_T(\bm{\theta}) & \bm{A}_T(\bm{\theta}) & \bm{D}_T(\bm{\theta})
            \end{bmatrix},\\
            \bm{R} &= \begin{bmatrix}
                \bm{A}_R(\bm{\phi}) & \bm{A}_R(\bm{\phi}) & \bm{A}_R(\bm{\phi}) & \bm{A}_R(\bm{\phi}) & \bm{D}_R(\bm{\phi}) & \bm{A}_R(\bm{\phi})
            \end{bmatrix},\\
            \bm{A}_T(\bm{\theta}) &= \begin{bmatrix}
                \bm{a}_T(\theta_1) & \bm{a}_T(\theta_2) & \dots & \bm{a}_T(\theta_L)
            \end{bmatrix},\qquad
            \bm{D}_T(\bm{\theta}) = \begin{bmatrix}
                \bm{d}_T(\theta_1) & \bm{d}_T(\theta_2) & \dots & \bm{d}_T(\theta_L)
            \end{bmatrix},\\
            \bm{A}_R(\bm{\phi}) &= \begin{bmatrix}
                \bm{a}_R(\phi_1) & \bm{a}_R(\phi_2) & \dots & \bm{a}_R(\phi_L)
            \end{bmatrix},\qquad
            \bm{D}_R(\bm{\phi}) = \begin{bmatrix}
                \bm{d}_R(\phi_1) & \bm{d}_R(\phi_2) & \dots & \bm{d}_R(\phi_L)
            \end{bmatrix}.
            \end{split}
            \label{eq:fimparam}
        \end{equation}
        \hrulefill
    \end{figure*}


\bibliographystyle{IEEEtran}
\bibliography{IEEEabrv,mybib.bib}
\end{document}